\documentclass[pre,aps,showpacs,preprintnumbers,amsmath,amssymb,superscriptaddress,twocolumn]{revtex4}
\usepackage{graphicx}
\usepackage{pstricks}
\usepackage{psfrag}

\newcommand{\vv}[1]{\mathbf{#1}}

\begin{document}

\title{Demography-based adaptive network model reproduces the spatial organization of human linguistic groups}

\author{Jos\'e A. Capit\'an}
\affiliation{Centro de Estudios Avanzados de Blanes (CSIC). C/ d'acc\'es a la Cala St. Francesc 14, 
17300 Blanes, Girona, Spain}
\author{Susanna Manrubia}
\affiliation{Centro Nacional de Biotecnolog\'{\i}a (CSIC). C/ Darwin 3, 28049 Madrid, Spain}

\date{\today}

\begin{abstract}
The distribution of human linguistic groups presents a number of interesting and non-trivial patterns. The distributions 
of the number of speakers per language and the area each group covers follow log-normal distributions, while population 
and area fulfill an allometric relationship. The topology of networks of spatial contacts between different linguistic 
groups has been recently characterized, showing atypical properties of the degree distribution and clustering, among 
others. Human demography, spatial conflicts, and the construction of networks of contacts between linguistic groups are 
mutually dependent processes. Here we introduce an adaptive network model that takes all of them into account 
and successfully reproduces, using only four model parameters, not only those features of linguistic groups already 
described in the literature, but also correlations between demographic and topological properties uncovered in this work. 
Besides their relevance when modeling and understanding processes related to human biogeography, our 
adaptive network model admits a number of generalizations that broaden its scope and make it suitable to represent 
interactions between agents based on population dynamics and competition for space. 
\end{abstract}



\maketitle

\section{Introduction}

Adaptive networks, where the dynamics of nodes is coupled to the dynamics of network links, have received 
considerable attention in the last decade~\cite{Gro09}. This kind of networks represents not only a natural extension 
of models where either dynamics on complex networks or the origin of non-trivial topology of networks itself had 
been the focus of attention, but is in its own right a field of interest. Indeed, in many natural systems node dynamics 
and network dynamics are intimately coupled, and their interplay captures important aspects that would be missed if 
both processes are not taken simultaneously into account. Examples are, among many others~\cite{Gro08}, neural 
networks, where neuron activity affects synaptic strength~\cite{Gjo11}, ecological networks, where population dynamics 
is coupled to food web structure~\cite{Dro01}, catalytic networks, where the appearance of auto-catalytic sets formed 
by sufficiently abundant chemical species is essential for the maintenance of the system~\cite{Jai01}, and where the 
explicit introduction of space leads to segregation of parasitic species that may otherwise disrupt network 
structure~\cite{Man03}, or generic models where distinct populations of nodes separate when connection strength is 
allowed to vary~\cite{Ito02}.

The coupling between node and link dynamics is especially relevant in social networks, where nodes are individuals, 
companies, human groups or countries, e.g., and links represent social contacts of various kinds. In many of these 
networks, agents can actively change their interactions, thus causing a systematic modification of network topology. 
A well studied case is that of epidemics, where susceptible individuals may suppress their links with infected neighbors, 
leading to networks assortative in degree and to first-order transitions between healthy and endemic states~\cite{Gro06} 
and even to infection suppression~\cite{Zan08b}. Similar situations hold in socioeconomic contexts, where adaptive 
networks display an interesting phenomenology that includes phase transitions and hysteresis between dissimilar 
states of agents~\cite{Ehr06} and self-organization leading to broad wealth distributions~\cite{Bur08}. In a broader 
scenario, it has been shown that changes in the state of nodes coupled to rewiring of links systematically causes 
network fragmentation~\cite{Her11}. This fact seems to be enhanced by the spatial embedding of many social 
networks, which constraints interactions~\cite{Erb14} and may induce the spatial separation of different socioeconomic 
classes~\cite{Met05}. 

In this work we address the relationship between the demographic dynamics of human linguistic groups and the 
topology of their networks of spatial contacts. We present a model for an adaptive spatial network where neighboring 
relationships are determined by the growth of groups and their concomitant attempt to modify the total area they 
occupy. The model is based on two previous and independent observations regarding the organization of human 
groups. In~\cite{Man12}, a mean-field model was introduced to reproduce the population-area relationship observed 
in human languages; the spatial structure of groups did not play any essential role in explaining that observation, and thus 
was disregarded. As a consequence, that model cannot describe the complex topology of the network of contacts 
that was later uncovered~\cite{Cap14}. Networks of contacts between linguistic groups reflect their spatial embedding 
and display a set of properties previously unseen in spatial networks. Among others, those networks have high 
intervality, a property shared with food webs~\cite{Sto06,Cap13}. Inspired in this latter system, and in niche models which 
had successfully captured that property, a niche-like algorithm was proposed and shown to recover most topological 
features of language networks~\cite{Cap14}. 
 
Human demographic dynamics and the spread of populations on space, which determines their inter-group contacts, 
are two coupled processes. As we report in this contribution, their mutual dependence is behind observed correlations 
between the population of a group and the number of spatial neighbors it has, and is needed to explain the appearance 
of assortative properties in empirical language networks. Results from an adaptive network model that we here introduce 
are compared with several world regions and to the topology of the network of linguistic contacts in each of them. The 
paper is structured as follows. In Section II we introduce the adaptive model for language networks. Previous relevant 
results are summarized for the sake of completeness and clarity. Section III reviews data on human linguistic groups as 
used in this study, particularly emphasizing the meaning of model parameters. In Section IV, we fit the model to empirical 
data and show how the adaptive model qualitatively and quantitatively, in most cases, reproduces the population-area 
relationship, the degree and shortest path distributions of empirical language networks (in addition to other topological 
features), and non-trivial correlations between demography and topology. The paper finishes with an overall discussion 
and some proposals for model extensions and future research.

\section{Adaptive model for language networks}

The adaptive network model yields a dynamic network of interactions among groups arising from explicit demographic 
dynamics and competition for space. As the size of groups varies, their neighboring relationships are modified and 
possible conflicts with different groups sharing boundaries may ensue. The precise example used is the development 
of human linguistic groups in the last thousand years. First, we define demographic dynamics following current knowledge 
on the world population growth and suitable rules for inter-group contacts and conflicts. Second, the network of 
contacts between groups is updated in the light of changes in the areas they occupy. 

\subsection{Demography and conflict rules}

The modeling of demographic dynamics is based on~\cite{Man12}. Dynamics relies on a stochastic multiplicative 
process of the form $P_i (t+i) = \alpha_i (t) P_i (t)$ for the size of each population $P_i$, 
where the distribution of $\alpha_i$ values is estimated from empirical 
data. This process describes the growth of linguistic groups~\cite{Zan08} and reproduces the observation of a log-normal 
distribution of the number of speakers per language~\cite{Sut03}. Subsequently, demographic changes are coupled to 
variations in the area over which groups are spread. That model was devised with the goal of explaining the population 
($P$)-area ($A$) relationship observed in human linguistic groups, which follows $A \propto P^z$~\cite{Man12}. 

Relevant model parameters have been derived from world population estimations, as follows. There were about 
$P_0 = 3.1 \times 10^8$ humans in year 1000~\cite{UN}, while in year 2000 the world population reached 
$P_T = 5.7 \times 10^9$~\cite{Gor05}. Assuming an exponential growth in the last ten centuries, an average annual 
growth rate $\alpha \simeq 1.0029$ is obtained, and a dispersion $\sigma_{\alpha}=0.096$ can be associated to the 
process~\cite{Zan08,note}. The simplest distribution for the stochastic growth rate $\alpha_i$ is a uniform distribution of 
average $\alpha$ and mean square dispersion $\sigma_\alpha$~\cite{Man12,Zan08}. A constant number of languages in 
this time interval, equal to the current estimated linguistic diversity (6900 languages)~\cite{Gor05}, is considered. 
Though some languages may have appeared in the last millenium, and many others have disappeared, in this model we 
disregard language birth or extinction for the sake of simplicity. In a previous model that constitutes the basis for the 
demographic dynamics here implemented, it has been numerically shown that those two processes did not affect the 
statistical results~\cite{Zan08}. As initial condition, we take uniform populations ($P_i(0)=3.1\times 10^8/6900$), and areas 
$A_i=1$, in arbitrary units. Numerical simulations show that changes in the initial condition do not affect in a significant way 
the final distribution of group sizes (see also~\cite{note}). Dynamics are run for $1000$ time steps to compare with current 
available data. In the scenario described, population dynamics are defined so as to agree with empirical observations. 
Therefore, parameters $\alpha$ and $\sigma_{\alpha}$ implicitly contain information on all processes that may have potentially 
affected demographic changes in the last millenium (that is births and deaths, but also casualties due to wars or pandemic 
diseases, for example). This is also the reason to couple in a directed fashion population dynamics to areas. Notice that the 
units of area remain undetermined to a multiplicative factor. 

The log-normal distributions of language sizes $P_i$~\cite{Sut03,Zan08} and areas $A_i$~\cite{Man12} imply that the 
log-transformed variables $p_i = \ln P_i$ and $a_i = \ln A_i$ for each linguistic group $i$ follow Gaussian distributions. 
As a result, the stochastic multiplicative process in the original variables can be cast in the form of a stochastic additive 
process in $p_i$ and $a_i$~\cite{Man12}. The logarithmic number of individuals in a group therefore follows
\begin{equation}
p_i(t+1) = p_i(t) + \beta_i(t) \, ,
\end{equation}
where $t$ is measured in years, and $\beta_i$ is randomly drawn at each time step from a uniform distribution 
$\Pi(\beta;\epsilon,\eta)$ in the interval $(\epsilon-\eta, \epsilon+\eta)$,
\begin{equation}
\Pi(\beta;\epsilon,\eta)=\frac{1}{2\eta} \left[\Theta(\beta-(\epsilon-\eta))-\Theta((\epsilon+\eta)-\beta) \right] \, ,
\end{equation}
with mean value $\epsilon=-0.00186$ and half-width $\eta=0.169$ obtained when the original multiplicative process is 
mapped to an additive one~\cite{Man12,note1}. Similarly, the logarithmic area $a_i$ is assumed to obey
\begin{equation}
a_i(t+1)=a_i(t)+\xi_i(t) \, .
\end{equation}
The evolution of $p_i$ and $a_i$ is coupled following two rules:
\begin{enumerate}
\item The area covered by a group shrinks when its population decreases: if $\beta_i(t)<0$, then $\xi_i(t)$ is randomly 
drawn from a uniform interval $[-r|\beta_i(t)|,0]$.
\item Increases in the population size lead to conflict between group $i$ and one of its neighbors on the network of 
contacts between groups (see below). A neighbor $j$ of node $i$ is chosen at random; if its growth rate is smaller than 
that of $i$, the area of $i$ grows, and {\it vice versa}. 
Specifically,
\begin{enumerate}
\item[(a)] If $\beta_j(t) < \beta_i(t)$, $\xi_i(t)$ is drawn from $[0,w\beta_i(t)]$;
\item[(b)] If $\beta_j(t) \ge \beta_i(t)$, $\xi_i(t)$ is drawn from $[-w\beta_i(t),0]$.
\end{enumerate}
\end{enumerate}
The {\em spontaneous retreat} parameter $r$ measures to which extent log-areas spontaneously shrink when 
populations decrease. The {\em outcome of conflicts} is weighted through $w$, which determines the associated 
benefit for the population with the faster growth and is, in general, different from $r$. Actually, mean-field fits to actual 
values of $p_i$ and $a_i$ at present have revealed a sub-linear relationship between population decrease and area 
reduction and a larger increase in areas as a result of conflicts, yielding $w>1>r$ for the six world regions analyzed 
in~\cite{Man12}. It remains to be seen whether this constraint remains in other world regions here analyzed, and 
whether the values of $w$ and $r$ in the current adaptive network model significantly deviate from mean-field results. 

\subsection{Network dynamics}
Space is effectively introduced in the form of a network of neighbors that coevolves with the demographic dynamics 
just described. The construction of the network is inspired in a static algorithm that used a given distribution of areas 
and contained the rules to construct a network of contacts between groups~\cite{Cap14}. Now, instead, the network is 
continuously updated taking into consideration the area associated to each node, as obtained in the previous step. 
Neighboring relationships between nodes arise from an assumption on perimeter contact. Based on geometric constraints, 
it can be assumed that the perimeter of node $i$ is comparable to the sum of perimeters of its potential neighbors up to a 
multiplicative factor, 
\begin{equation}\label{eq:perimeter}
A_i^{1/2}\simeq f_i \sum_{j\in\mathrm{nn}(i)} A_j^{1/2} \, ,
\end{equation}
where the {\em perimeter overlap} $f_i > 0$ measures the average fraction of perimeter of each neighbor that is shared 
with node $i$. In general, $f_i$ ---as defined in Eq.~\eqref{eq:perimeter}--- is a node-dependent quantity, but for simplicity 
we assume an effective value all across the network, such that $f_i$ will be substituted by its network average 
$f = N^{-1} \sum_i f_i$, where $N$ is the number of nodes (languages) in the network.

The network of contacts is generated in two steps:
\begin{enumerate}
\item \emph{Directed network generation.} Given the (arbitrarily ordered) set of areas $\{A_1(t),A_2(t),\dots,A_N(t)\}$ at 
time step $t$, where $A_i(t)=e^{a_i(t)}$, we draw \emph{directed} links between each node $i$ and nodes at positions 
$i\pm 1,i\pm 2,\dots$, until the upper bound of the rhs of Eq.~\eqref{eq:perimeter} is first exceeded. The first neighbor 
$j_0$ is either $i+1$ or $i-1$ with equal probability, and subsequent nodes are chosen following the rules
\begin{equation}
j_{2n+1}=\begin{cases}
i-n-1, & j_0=i+1,\\
i+n+1, & j_0=i-1,\\  
\end{cases}
\end{equation}
and
\begin{equation}
j_{2n}=\begin{cases}
i+n+1, & j_0=i+1,\\
i-n-1, & j_0=i-1,\\  
\end{cases}
\end{equation}
for $n=0,1,2,\dots$ Periodic boundary conditions have been assumed when $j_n < 0$ or $j_n > N$.

\item \emph{Transformation to an undirected network.} Since spatial neighboring relationships are undirected, the 
previous network should be transformed to an undirected one. Links may be added or removed so as to guarantee the 
symmetry of the adjacency matrix. For this purpose we introduce a {\em symmetrization parameter} $0\le q\le 1$. If a 
directed link $i\to j$ does not have a reverse counterpart, $j\not\to i$, we draw a uniformly distributed random number 
$x$ in $(0,1)$ and add the missing link $j\to i$ to the network if $x<q$. If $x\ge q$, the original link $i\to j$ is removed. 
In any case, the relationship between $i$ and $j$ has been symmetrized after the process. Note that this process 
affects neighboring relationships as defined in Eq.~\eqref{eq:perimeter}, so it will be important to assess its eventual 
effect in the demographic and topological properties we aim at reproducing. 
\end{enumerate}

\begin{figure}[!t] 
 \centering 
\includegraphics[width=0.48\textwidth]{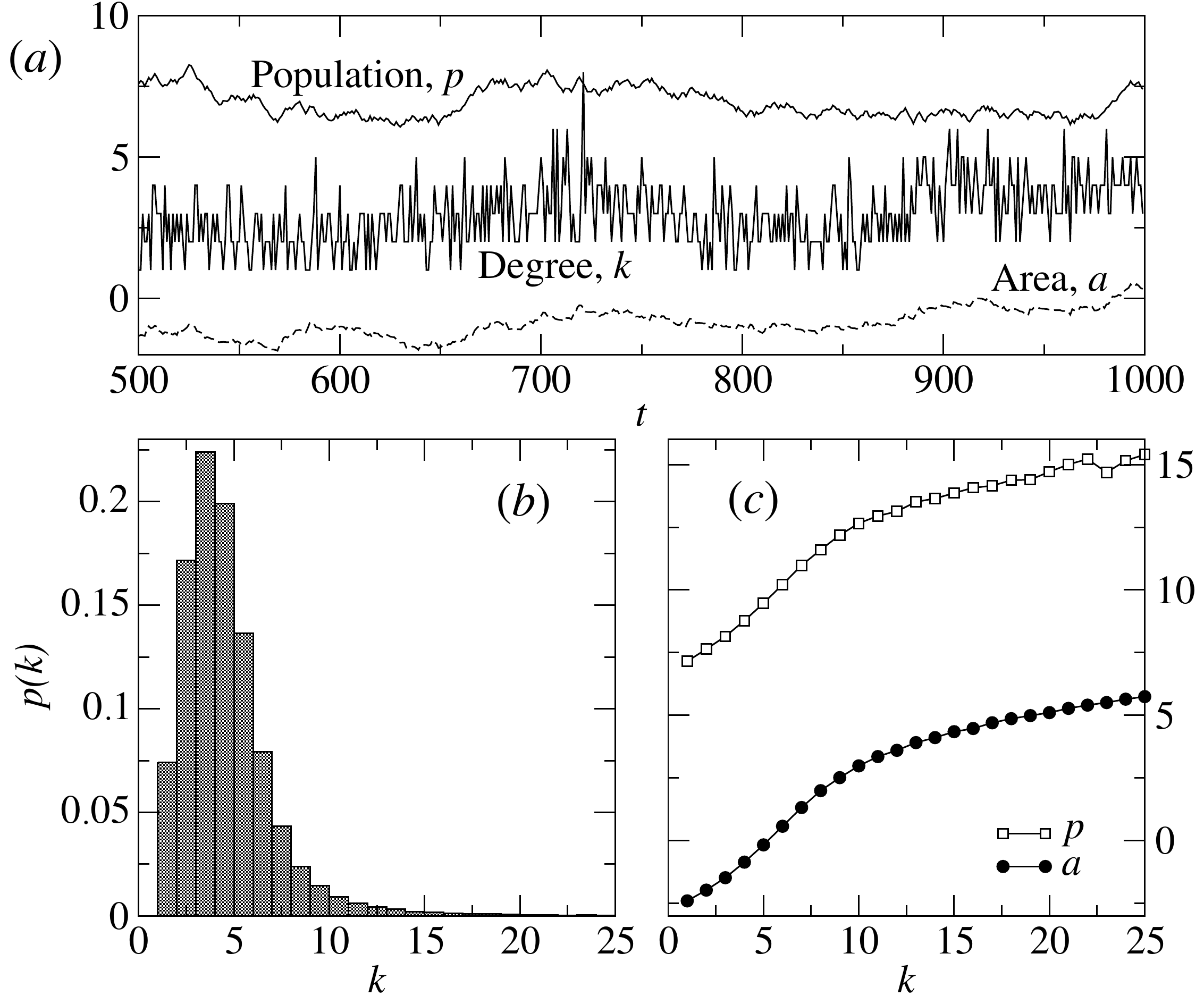}
\caption{Model dynamics. (a) Time series corresponding to a realization of the model showing the dynamics of the 
natural logarithm of area and population size for a single linguistic group, and the number of neighboring languages it 
has. (b) Degree distribution obtained at the end of the realization for a system with 1000 interacting groups. (c) 
Relationship between the number of neighbors $k$ and the total area or population for the same ensemble. Parameters 
are $r=1$, $w=1.5$, $f=0.2$, and $q=0.3$, and averages over $100$ and $500$ independent realizations 
have been performed in (b) and (c), respectively.}
\label{fig:DynModel}
\end{figure}

The use of a one-dimensional array of areas to construct the network is analogous to the procedure used in ecological 
niche models, where a single variable suffices to reproduce most topological properties of food webs and where the 
explicit consideration of population dynamics is not essential. In the case of networks of contacts between linguistic 
groups, their local structure was shown to be equivalent to that of almost regular, one-dimensional networks, with the 
area playing the role of the niche variable~\cite{Cap14}. 

Figure~\ref{fig:DynModel} illustrates some important properties of the model just described. Fig.~\ref{fig:DynModel}(a) 
exemplifies the dynamics of logarithmic areas $a_i$ and populations $p_i$, as well as the number of neighbors of group 
$i$ --its degree $k_i$-- for 500 years. At the end of the simulation [Fig.~\ref{fig:DynModel}(b)], for $t=1000$, the degree 
distribution $p(k)$ is calculated. It presents a well-defined average value and a significant tail to large $k$ values. Finally, 
in Fig.~\ref{fig:DynModel}(c) we illustrate one of the quantities that could not be reproduced by models based only on 
demography~\cite{Man12} or on a niche-like algorithm to construct the network~\cite{Cap14}, namely, the relationship 
between population or area and degree of a linguistic group. Among others, these quantities will be compared to those 
measured in current linguistic groups with the purpose of establishing whether the dynamical niche model is able to 
reproduce the observations and of determining the value of the model parameters that best fits the latter. Summarizing, 
the dynamical niche model is characterized by four parameters: the spontaneous retreat $r$, the outcome of 
conflicts $w$, the average perimeter overlap $f$, and the symmetrization parameter $q$. 

\section{Data on human linguistic groups}

Data on linguistic groups stems from a collection by SIL International (http://www.ethnologue.com/) and a map developed 
by Global Mapping International (world LanguageMapping System, http://www.gmi.org/wlms/index.htm). A detailed 
description of the database appears in the Ethnologue~\cite{Gor05}, from which information on 6900 extant languages, 
including their spatial distribution and number of speakers, can be found. Each language is characterized by a centroid, 
which is a point in latitude-longitude coordinates that represents its average location. Centroids are the nodes of linguistic 
groups. Two nodes are linked if the groups they represent share spatial borders in any of the domains where a language is 
spoken (note that the speakers of a language may occupy disconnected domains, a situation that is relatively frequent). 
The interested reader can find details on network construction in~\cite{Cap14}. 

In the present study, we are not taking into account languages which are widespread as a result of colonization. 
Languages such as English, Spanish or Portuguese in the Americas, or Mandarin Chinese in Asia, are in several senses 
outsiders: they percolate across continental regions and act as hubs in networks of contacts between linguistic groups, 
enhancing the formation of large connected components in language networks~\cite{Cap14}. The number of neighbors of 
widespread languages (that is, their degree $k_i$) is several-fold higher than that of other languages in the same network, 
thus significantly deviating from the bulk degree distribution. In this sense, widespread languages can be considered the 
Dragon Kings of languages~\cite{Sor09}, and the dynamical processes that underlie their spread are different from the 
basic demographic dynamics implemented in our model. Widespread languages constitute a small fraction of world 
languages. The $50$ largest languages (with $24$ or more million speakers) represent only about $0.7\%$ of the data 
points here considered. The elimination of widespread languages causes the fragmentation 
of otherwise connected networks in some world regions, remarkably in continental North America. This effect is not seen 
if, for instance, the largest languages are eliminated in the network corresponding to continental Africa, whose largest 
connected component remains essentially unchanged.  

The main properties of 12 networks of linguistic groups selected for the current study are reported in Table~\ref{tab:data}. 
They correspond to five continental regions (Africa, Asia, Europe, and North and South America), though in the case of North 
America no large connected component can be identified: the three largest networks are found in or around Mexico, and 
named Mex1, Mex2 and Yucatan. In addition, we study the networks of Australia, New Guinea, Sulawesi and Luzon islands, 
as well as an additional small network found in the borders shared by Argentina, Bolivia, and Paraguay (ABP). 

\begin{table}[t!]
\begin{center}
\begin{tabular}{lcccccc}
\hline\hline
CC & $N$ & $L$ & $z$ & $\rho$ & $f$ & $\sigma_f$ \\
\hline\hline
C Africa & 2126 & 6154 & 0.87 & 0.63 & 0.13 & 0.11 \\
\hline
C Asia & 1370 & 3967 & 0.65 & 0.72 & 0.10 & 0.10 \\
New Guinea & 663 & 1543 & 0.62 & 0.42 & 0.22 & 0.16 \\
Australia & 99 & 176 & 0.72 & 0.52 & 0.28 & 0.15 \\
Sulawesi & 64 & 121 & 0.66 & 0.77 & 0.25 & 0.19 \\
Luzon & 56 & 140 & 0.44 & 0.79 & 0.13 & 0.08 \\
\hline
C Europe & 231 & 547 & 0.60 & 0.65 & 0.13 & 0.31 \\
\hline
Mex1 & 68 & 120 & 0.82 & 0.59 & 0.32 & 0.24 \\
Yucatan & 50 & 111 & 1.22 & 0.61 & 0.27 & 0.57 \\
Mex2 & 39 & 71 & 0.67 & 0.73 & 0.32 & 0.24 \\
\hline
CS America & 234 & 399 & 0.40 & 0.56 & 0.28 & 0.17 \\
ABP & 33 & 59 & 0.66 & 0.76 & 0.37 & 0.42 \\
\hline\hline
\end{tabular}
\end{center}
\caption{
Largest connected components of networks of contacts between linguistic groups obtained for each continent. The 
number of nodes $N$ and the number of links $L$ are shown. Relevant quantities that the model intends to reproduce 
are the exponent $z$, the correlation $\rho$, and the average perimeter overlap $f$. The deviation of the distribution of 
$f_i$ values in each network is $\sigma_f$. CC: Connected component; C Africa: continental Africa; C Asia: continental 
Asia; New Guinea, Sulawesi and Luzon are islands. C Europe: continental Europe; Mex1: Mexico (1); Yucatan: 
Yucatan peninsula; Mex2: Mexico (2); CS America: continental South America; ABP: ABP borders.
}
\label{tab:data}
\end{table}

\begin{figure}[!t] 
\centering 
\includegraphics[width=0.48\textwidth]{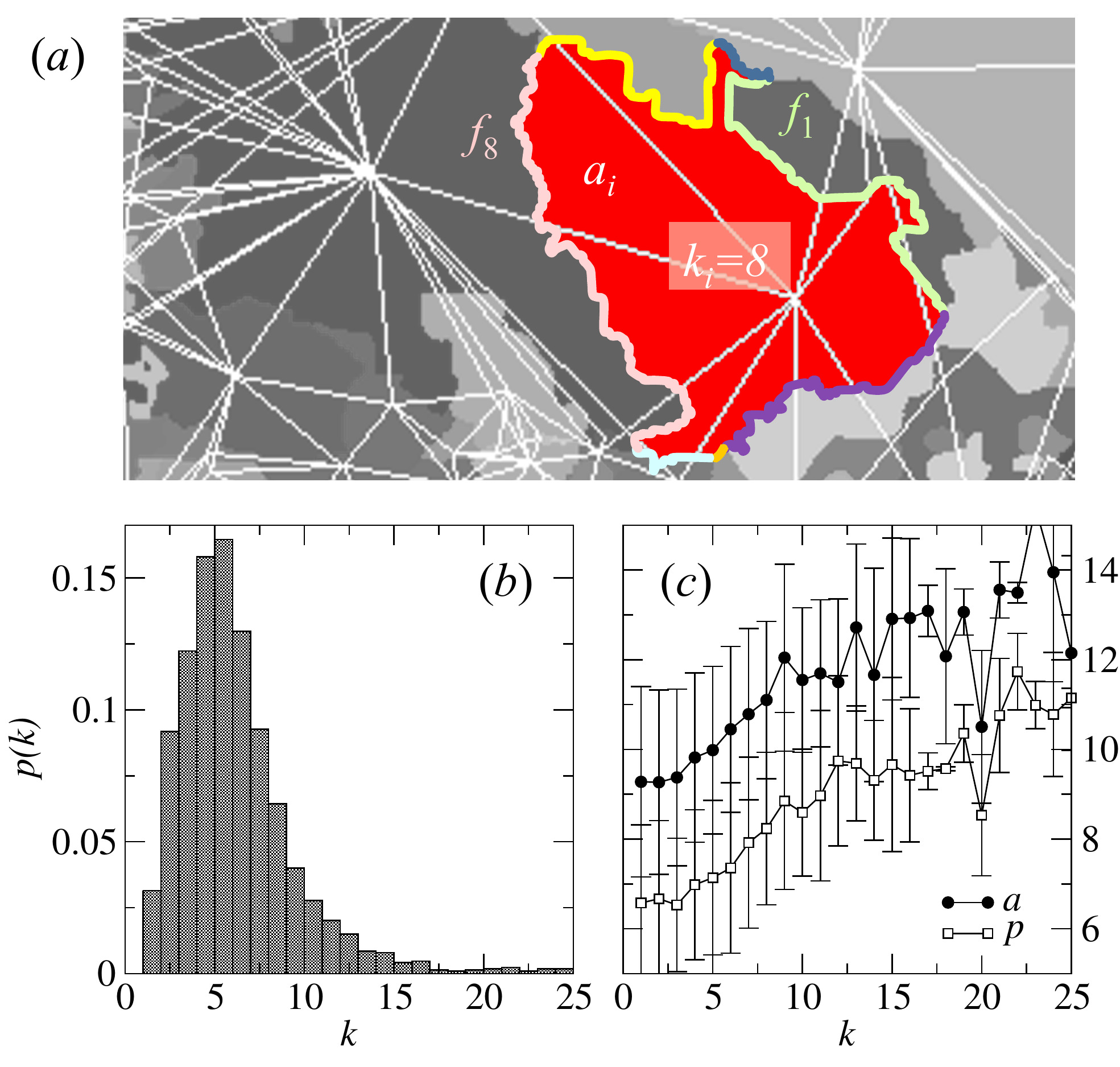}
\caption{Properties of African largest connected component, with $2126$ nodes. (a) Detail of the local structure of 
contacts between groups. Grey regions represent different domains where a language is spoken; one language might be 
spoken in disconnected domains. The centroid corresponds to the whole of a language, and therefore might fall even outside 
a particular domain; (b) Degree distribution; (c) Dependence of the logarithmic area $a$ and population $p$ of 
linguistic groups on the number of neighbors $k$ in the network of spatial contacts. Error bars stand for the standard 
deviation of $a$ and $p$ values at each fixed $k$.}
\label{fig:Africa}
\end{figure}

An example of some model quantities and empirical properties of the continental Africa network are represented in 
Fig.~\ref{fig:Africa}. In Fig.~\ref{fig:Africa}(a) a part of the whole network is shown, emphasizing the area $A_i$ of a 
given linguistic domain $i$ and its neighboring relationships. As can be seen, the perimeter overlap depends on each 
pair of groups in contact. In this example, language $i$ shares boundaries with eight different languages, so its has a 
degree $k_i=8$. In practice, $f_i$ is calculated from its definition, $f_i \simeq A_i^{1/2} / \sum_{j \in \rm{nn}(i)} A_j^{1/2}$ 
for each node, and then averaged over the whole network to obtain the value $f$ reported in Table~\ref{tab:data}. For 
completeness, the last column of Table~\ref{tab:data} summarizes, for each network in the dataset, the standard 
deviation $\sigma_f$ of the distribution of $f_i$ values. 

\subsection{Population-area relationship}

The relationship between the logarithm of the size of a linguistic group and the logarithm of the area over which its 
speakers are spread follows an allometric relationship that has a counterpart in ecology, where the abundance of a 
species and its home range are similarly related~\cite{Har01,Gas02}. It has been shown that area $a$ and population 
$p$ fulfill $a = z p + c$ for the whole world, where $c$ is a constant, for six large continental regions, and also for 
groups of hunter-gatherers~\cite{Man12}. Since language sizes and areas follow log-normal distributions, the transformed 
logarithmic variables $a$ and $p$ are well fitted by Gaussian distributions, and their joint distribution can be approximated 
by a bivariate normal distribution. This joint distribution is characterized by $z$, which is the slope of the major ellipse 
axis of the scatter plot containing all $(p_i,a_i)$ pairs, and a coefficient $\rho$ that quantifies how correlated $a$ and 
$p$ are.

Let us define, for each network, the average logarithmic area $\langle a \rangle$ and the average logarithmic population 
$\langle p \rangle$,
\begin{equation}
{\langle a \rangle} = N^{-1} \sum_i a_i \, , \,\,\,\,\, {\langle p \rangle} = N^{-1} \sum_i p_i \, ,
\end{equation}
and the corresponding standard deviations
\begin{equation}
\sigma_a^2 = N^{-2} \sum_i (a_i-{\langle a \rangle})^2 \, , \,\,\,\,\, \sigma_p^2 = N^{-2} \sum_i (p_i-{\langle p \rangle})^2 \, .
\end{equation}
The covariance matrix $\mathsf{C}$ of $a$ and $p$ is 
\begin{equation}
\mathsf{C} = 
\left(
\begin{array}{cc}
\sigma_a^2 & \rho \sigma_a\sigma_p \\
\rho \sigma_a \sigma_p & \sigma_p^2
\end{array}
\right)
\end{equation}
with $\rho \sigma_a \sigma_p  = N^{-2} \sum_i (a_i -{\langle a \rangle}) (p_i - {\langle p \rangle})$. The eigenvectors of 
matrix $\mathsf{C}$ can be written as $(1,z)$, $(-z,1)$, where $z$ corresponds to the exponent relating both quantities. 
The value of $\rho$ determines the degree of correlation between $a$ and $p$: The larger $\rho$, the more correlated 
the two variables are. Values of $z$ and $\rho$ obtained through this procedure for the networks here analyzed are 
reported in Table~\ref{tab:data}. The interested reader can find example plots of this relationship for empirical data 
in~\cite{Man12}.

\subsection{Topological properties}

Networks of contacts between linguistic groups hold a number of non-trivial topological properties~\cite{Cap14}. They 
are an example of quasi-interval graphs, a property they share with food webs~\cite{Coh77,Sto06,Cap13}. The dependence 
between the clustering coefficient and the linkage density $2L/N$ reveals that language networks are akin to 
one-dimensional regular networks at the local level~\cite{Cap14}. Together with intervality, this property supports the 
existence of a configuration space of low dimensionality, and partly explains the success of a niche-like algorithm to account 
for several of the topological properties of language networks. Two additional properties that we will analyze and compare 
with model results are the shortest path length and the degree distributions. 

A representative example of the degree distribution $p(k)$ is shown in Fig.~\ref{fig:Africa}(b). Most language networks 
analyzed so far present degree distributions compatible with log-normal functions~\cite{Cap14}. In this work, one of our 
goals is to find out how likely it is that the adaptive network model generates degree distributions compatible with 
observations. The same applies to the distribution of shortest path lengths $p(d)$. The latter have a complex shape that 
depends on the particular network, as will be shown. We have chosen these two distributions because of their presumable 
relevance regarding inter-group dynamics. For example, the degree distribution is related to the likelihood of entering into 
conflict with different linguistic (or cultural) groups as a result of shared boundaries, but may also affect linguistic evolution 
due to frequent contacts with dissimilar languages. The shortest path length distribution may play a role in the 
dissemination of cultural innovations, under the reasonable assumption that intra-group spread of novelties is significantly 
faster than inter-group spread: the lesser intermediates, the faster the propagation. 

\subsection{Dependence between demographic and topological variables}

Demographic and topological features of language networks are not independent. For instance, Fig.~\ref{fig:Africa}(c) 
illustrates the empirical dependence of the logarithmic area $a$ and population $p$ on the degree $k$. This relation is 
qualitatively similar to the dependence yielded by the adaptive network model, see Fig.~\ref{fig:DynModel}. In forthcoming 
sections we will make this relation quantitative by optimizing the values of model parameters to fit empirical observations. 
There is a final observation as yet unexplained, which is the appearance of population-population, area-area and 
degree-degree correlations between neighboring nodes in empirical networks (see below).

\section{Model parameters: fits to empirical data}

With the aim of quantitatively reproducing demographic and topological features of language groups and networks, we 
analyze which values of the model parameters $r$, $w$, $f$, $q$ best fit each of the $12$ empirical networks considered. 
Specific goals are to reproduce the empirical parameters $z$ and $\rho$ characterizing the (logarithmic) population-area 
relationship, and two topological features: the degree distribution and the distribution of shortest-path lengths. We will finally 
evaluate how the dynamical model with the so obtained parameters reproduces the relationship between demographic
and topological variables, as well as the appearance of correlations in node properties (area, population, and degree). 

Quantitative values of $z$ and $\rho$ obtained with the adaptive network model fixing values of $f^{\star}$ and $q^{\star}$ 
do not differ substantially from those obtained in the mean-field approximation used in~\cite{Man12}, see Table~\ref{tab:rwfit}. 
The mean-field coupled dynamics of growth and conflict cannot be fitted to New Guinea island, contrary to what was 
observed in~\cite{Man12}. Such discrepancy is due to the fact that we are only considering here connected networks, 
disregarding isolated languages that were taken into account in~\cite{Man12} when calculating the empirical values of $z$ 
and $\rho$. Additionally, in that reference the mean-field dynamics could not be fitted to the pooled set of North American 
languages, for which the correlation value was sensibly smaller than the rest ($\rho=0.40$). Here the same phenomenon 
occurs for New Guinean connected languages ($\rho=0.42$, see Table~\ref{tab:data}). Consequently, the adaptive network 
model cannot be fitted to New Guinea network; hence, from now on, we will reduce our analysis to the remaining $11$
networks.

The relative difference between the surfaces $z(r,w,f^{\star},q^{\star})$ and $\rho(r,w,f^{\star},q^{\star})$ and their mean-field 
counterparts $z_{\mathrm{MF}}(r,w)$ and $\rho_{\mathrm{MF}}(r,w)$ has been measured as 
\begin{equation}
n_z=\max_{i,j} \left|\frac{z(r_i,w_j,f^{\star},q^{\star})-z_{\mathrm{MF}}(r_i,w_j)}{z_{\mathrm{MF}}(r_i,w_j)}\right|
\end{equation}
and
\begin{equation}
n_{\rho}=\max_{i,j} \left|\frac{\rho(r_i,w_j,f^{\star},q^{\star})-\rho_{\mathrm{MF}}(r_i,w_j)}{\rho_{\mathrm{MF}}(r_i,w_j)}\right|
\end{equation}
for discretizations $(r_i,w_j)$ of the $(r,w)$ parameter subspace, with $r_i\in (0,1.5)$, $w_j\in (0.5,3)$, $f^{\star}=0.1$
and $q^{\star}=0.1$. Comparison 
with mean field values yields $n_z=0.12$ and $n_{\rho}=0.019$, which implies that maximum relative differences are 
around $12\%$ and $2\%$ for $z$ and $\rho$, respectively, when network dynamics is explicitly considered. Since the 
mean-field model does not depend on $f$ and $q$, this result suggests a weak dependence of the population-area relationship (through variables $z$ and $\rho$) on the latter parameters. Deeper numerical explorations show that 
$\rho$ is almost independent of $f$ and $q$, whereas $z$ varies moderately in the region $q \approx 0$, becoming 
almost constant for $q > 0.3$.

Assuming that the parameter subspace $(r,w)$ is mostly uncoupled to the subspace $(f,q)$, our fits will be performed 
in two steps. First, we fit $(r,w)$ to empirical values of $z$ and $\rho$, keeping $f^{\star}=0.1$ and $q^{\star}=0.1$ fixed 
(see Table~\ref{tab:data}). Second, we obtain estimates for $f$ and $q$ by imposing that simulated degree and 
shortest-path length distributions keep close (in a precise sense to be defined) to empirical distributions. Finally, we check 
that the estimates of $z$ and $\rho$ still reproduce empirical values for the $f$ and $q$ obtained. 

\begin{table}[t!]
\begin{center}
\begin{tabular}{lcccc}
\hline\hline
CC & $r_{\rm MF}$ & $w_{\rm MF}$ & $r$ & $w$ \\
\hline\hline
C Africa & $0.651$ & $2.23$ & $0.72\pm 0.02$ & $2.31\pm 0.02$ \\
\hline 
C Asia & $0.951$ & $1.54$ & $1.00\pm 0.01$ & $1.56\pm 0.01$ \\
Australia & $0.125$ & $2.20$ & $0.16\pm 0.05$ & $2.27\pm 0.06$ \\
Sulawesi & $1.27$ & $1.25$ & $1.32\pm 0.07$ & $1.23\pm 0.08$\\
Luzon & $1.04$ & $0.735$ & $1.07\pm 0.14$ & $0.70\pm 0.16$ \\\hline
C Europe & $0.603$ & $1.69$ & $0.64\pm 0.02$ & $1.71\pm 0.02$ \\
\hline
Mex1 & $0.422$ & $2.24$ & $0.48\pm 0.07$ & $2.30\pm 0.07$ \\
Yucatan & $0.695$ & $2.79$ & $0.84\pm 0.12$ & $2.92\pm 0.13$ \\
Mex2 & $1.16$ & $1.16$ & $1.11\pm 0.08$ & $1.49\pm 0.09$ \\\hline
CS America & $0.195$ & $1.47$ & $0.20\pm 0.01$ & $1.48\pm 0.02$ \\
ABP & $1.20$ & $1.33$ & $1.27\pm 0.10$ & $1.28\pm 0.11$ \\
\hline\hline
\end{tabular}
\end{center}
\caption{
Optimal mean-field parameters (obtained with the model in~\cite{Man12}) and adaptive network model parameters $r$, 
$w$ for $f^{\star}=0.1$ and $q^{\star}=0.1$.}
\label{tab:rwfit}
\end{table}

\subsection{Fitting \emph{r} and \emph{w} to data}

As initial guesses for $(r,w)$, we use the mean-field values $(r_{\mathrm{MF}},w_{\mathrm{MF}})$ reported in 
Table~\ref{tab:rwfit}. Subsequently, the adaptive network model is simulated in square neighborhoods of 
$(r_{\mathrm{MF}},w_{\mathrm{MF}})$, keeping $f^{\star}=0.1$ and $q^{\star}=0.1$ fixed. We use model networks 
with the same sizes of empirical ones and average over $1000$ model realizations. For each point $(r_i,w_j)$ of the 
grid, we estimate averages over realizations $z_{ij}=\langle z\rangle(r_i,w_j)$ and $\rho_{ij}=\langle\rho\rangle(r_i,w_j)$, 
as well as the corresponding standard deviations $\sigma_{ij}^z$ and $\sigma_{ij}^{\rho}$. 

Let us define $\vv{y}=(z,\rho)^{\mathrm{T}}$ and $\vv{x}=(r,w)^{\mathrm{T}}$. In a local neighborhood of each point of 
the grid we expect an approximate (two-dimensional) linear dependence between $\vv{y}$ and $\vv{x}$, 
\begin{equation}
\vv{y}\approx \mathsf{M}\vv{x}+\vv{b}
\end{equation}
for a constant $2\times 2$ (Jacobian) matrix $\mathsf{M}=(m_{ij})$ and a vector $\vv{b}=(b_1,b_2)^{\mathrm{T}}$ to be 
determined. We estimate the required coefficients by means of a two-dimensional, weighted least-squares fit to simulated 
data, i.e.,
\begin{equation}
\begin{aligned}
z_{ij} &= m_{11}r_i+m_{12} w_j+b_1,\\
\rho_{ij} &= m_{21}r_i+m_{22} w_j+b_2.
\end{aligned}
\end{equation}
Fit's weights are chosen in the usual way, as $1/\sigma^2_{ij}$, provided that standard deviations for $z$ and $\rho$ are 
known. Note that the least-squares method provides estimates for standard errors of $m_{ij}$ and $b_i$. 

Finally, $r$ and $w$ estimates come from 
\begin{equation}\label{eq:sol}
\vv{x}=\mathsf{M}^{-1}(\vv{y}-\vv{b}).
\end{equation}
The errors of $r$ and $w$ have been calculated using standard error propagation according to Eq.~\eqref{eq:sol}. Results 
are listed in Table~\ref{tab:rwfit}, where they can be compared to mean-field estimates. There are some quantitative 
differences regarding previous results in different world regions~\cite{Man12}. The value of the spontaneous retreat $r$ is 
not always below $1$, implying that the reduction in area caused by a decrease in the population size is not sub-linear in 
all cases. The four exceptions coincide with the smallest networks in our data set (Sulawesi, Luzon, Mex2 and ABP), so it 
cannot be discarded that this effect reflects a limited statistical power. The relationship $r < w$ holds in most cases,
with the exception of Sulawesi and Luzon islands.  

\subsection{Fitting \emph{f} and \emph{q} to data}

Now we proceed with the fit of $f$ and $q$ to topological quantities. For each network's size, we simulate $2000$ model 
realizations keeping $r$ and $w$ fixed and equal to the estimates previously obtained. The estimation of $f$ and $q$ 
proceeds through two different approaches: (i) minimizing the separation between the empirical and the simulated degree 
distributions; (ii) jointly adjusting the degree and the shortest path length distributions.

\subsubsection{Optimization based on the degree distribution}

We determine $f$ and $q$ as the values that minimize the Hellinger distance between the empirical degree distribution 
and the simulated degree distribution. For two arbitrary discrete distributions $\vv{g}=(g_i)$ and $\vv{h}=(h_i)$, the 
Hellinger distance~\cite{Hel09} is defined as
\begin{equation}
d_{\mathrm{H}}(\vv{g},\vv{h})=\frac{1}{\sqrt{2}}\sum_i\left(\sqrt{g_i}-\sqrt{h_i}\right)^2
=\frac{1}{\sqrt{2}}\parallel\sqrt{\vv{g}}-\sqrt{\vv{h}}\parallel_2,
\end{equation}
i.e., $d_{\mathrm{H}}$ is proportional to the Euclidean norm of the difference of square-root vectors. We choose the pair 
$(f,q)$ that minimizes $d_{\mathrm{H}}$ for all networks here considered, where $g_k=p_{\mathrm{e}}(k)$ is the 
empirical degree distribution, and $h_k=p_{\mathrm{s}}(k)$ is the simulated degree distribution. 

Minimization has been carried out in two steps: first we perform a parameter screening in $f\in[0.05,1]$ and $q\in[0,1]$. 
This yields an estimate of the pair $(f,q)$ that minimizes $d_{\mathrm{H}}$. Second, we use the estimation as initial 
guess for a standard algorithm of numerical minimization. 

For large values of $f$ and small values of $q$, the adaptive network model yields disconnected graphs. This is due to the 
fact that large $f$ values imply a small number of neighbors (therefore a low connectivity), while small $q$ values tend to 
eliminate all unpaired links. Therefore, the number of nodes of the giant component can be well below the size of the 
empirical network.
In order for the sizes of empirical and model networks to be comparable, the range of $f$ and $q$ values needs to be 
restricted. We require
\begin{equation}\label{eq:avN}
\langle N\rangle \ge 0.9N_{\mathrm{e}},
\end{equation}
$N_{\mathrm{e}}$ being the empirical network size. Parameter values yielding networks sizes outside of this region 
are not taken into account during minimization. This process yields the estimates listed in Table~\ref{tab:fqfitdeg} for 
$f$ and $q$, and Hellinger's distance for the degree and the shortest-path length distributions. 

\begin{table}[t!]
\begin{center}
\begin{tabular}{lccccc}
\hline\hline
CC & $f$ & $q$ &  $d_{\mathrm{H}}$ (degree) & $d_{\mathrm{H}}$ (path)\\
\hline\hline
C Africa & $0.14\pm 0.01$ & $0.30\pm 0.01$ & $0.09$ & $0.16$\\
\hline
C Asia & $0.15\pm 0.01$ & $0.30\pm 0.01$ & $0.12$ & $0.56$ \\
Australia & $0.12\pm 0.01$ & $0.05\pm 0.01$  & $0.08$ & $0.17$ \\
Sulawesi & $0.20\pm 0.01$ & $0.20\pm 0.01$  & $0.14$ & $0.17$ \\
Luzon & $0.11\pm 0.01$ & $0.00\pm0.01$  & $0.13$ & $0.44$ \\
\hline
C Europe & $0.15\pm 0.01$ & $0.18\pm 0.02$  & $0.15$ & $0.25$ \\
\hline
Mex1 & $0.14\pm 0.01$ & $0.07\pm 0.01$  & $0.15$ & $0.37$ \\
Yucatan & $0.46\pm 0.01$ & $0.71\pm 0.01$  & $0.21$ & $0.23$ \\
Mex2 & $0.49\pm 0.02$ & $0.84\pm 0.03$  & $0.24$ & $0.19$ \\
\hline
CS America & $0.17\pm 0.01$ & $0.13\pm 0.01$  & $0.12$ & $0.16$ \\
ABP & $0.15\pm 0.01$ & $0.12\pm 0.01$ & $0.14$ & $0.10$ \\
\hline\hline
\end{tabular}
\end{center}
\caption{
Fitted model parameters $f$ and $q$ obtained through minimization of the distance between simulated and empirical 
degree distributions.}
\label{tab:fqfitdeg}
\end{table}

\subsubsection{Optimization based on the degree and shortest path length distributions}

We now apply a joint minimization of Hellinger's distance to the degree and shortest path length distributions. The sum 
of both distances is used as the objective function to minimize, 
\begin{equation}
s(f,q)=d_{\mathrm{H}}(\mathbf{d},\mathbf{e})+d_{\mathrm{H}}(\mathbf{g},\mathbf{h}),
\end{equation}
where $\mathbf{d}=(p_{\mathrm{e}}(d))$ is the empirical distribution of shortest-path length and 
$\mathbf{e}=(p_{\mathrm{s}}(d))$ is its simulated counterpart, and similarly for the empirical and simulated
 degree distributions $\mathbf{g}=(p_{\mathrm{e}}(k))$ and $\mathbf{h}=(p_{\mathrm{s}}(k))$. 
 Results are listed in Table~\ref{tab:fqfitjoint}. The restriction given by Eq.~\eqref{eq:avN} also applies here.

\begin{table}[t!]
\begin{center}
\begin{tabular}{lccccc}
\hline\hline
CC & $f$ & $q$ &  $d_{\mathrm{H}}$ (degree) & $d_{\mathrm{H}}$ (path) \\
\hline\hline
C Africa & $0.11\pm 0.01$ & $0.14\pm 0.01$ & $0.13$ & $0.11$\\
\hline
C Asia & $0.09\pm 0.01$ & $0.25\pm 0.01$ & $0.28$ & $0.06$ \\
Australia & $0.11\pm 0.01$ & $0.01\pm 0.01$  & $0.10$ & $0.07$\\
Sulawesi & $0.21\pm 0.01$ & $0.20\pm 0.02$  & $0.15$ & $0.14$ \\
Luzon & $0.13\pm 0.01$ & $0.29\pm0.06$  & $0.31$ & $0.15$ \\
\hline
C Europe & $0.11\pm 0.01$ & $0.12\pm 0.01$  & $0.19$ & $0.04$\\
\hline
Mex1 & $0.67\pm 0.03$ & $0.85\pm 0.03$ & $0.18$ & $0.19$\\
Yucatan & $0.11\pm 0.03$ & $0.06\pm 0.04$  & $0.26$ & $0.08$ \\
Mex2 & $0.61\pm 0.01$ & $0.97\pm 0.01$  & $0.26$ & $0.15$ \\
\hline
CS America & $0.16\pm 0.01$ & $0.12\pm 0.02$  & $0.13$ & $0.11$ \\
ABP & $0.16\pm 0.01$ & $0.11\pm 0.01$ & $0.15$ & $0.06$\\
\hline\hline
\end{tabular}
\end{center}
\caption{
Fitted model parameters $f$ and $q$ using the joint minimization scheme.}
\label{tab:fqfitjoint}
\end{table}

Note that too large values of $f$ or too small values of $q$ might cause a transition from a large connected component 
to a mostly disconnected ensemble of small networks when the network dynamics step is applied. The effect of 
decreasing $f$ and/or increasing $q$ from sufficiently low values (where nodes are disconnected) eventually causes a 
percolation transition comparable to that described in the standard Erd\H{o}s-Renyi model~\cite{Erd60} as the number of 
links increases. The fitted values we have obtained seem to balance such that the resulting networks are connected. For 
example, in Table~\ref{tab:fqfitdeg} we observe that low $f$ values correspond to mostly low $q$ values (as in Luzon, 
Australia, continental Africa and Mex1), while high $f$ values are associated to high $q$ values (as in Yucatan and Mex2). 
This association is also observed in Table~\ref{tab:fqfitjoint}, with interesting, consistent inversions of the correspondence 
seen in Mex1 and Yucatan. 

\subsection{Performance of the model}

\subsubsection{Degree distributions}

\begin{figure}[t!]
\begin{center}
\includegraphics[width=0.48\textwidth]{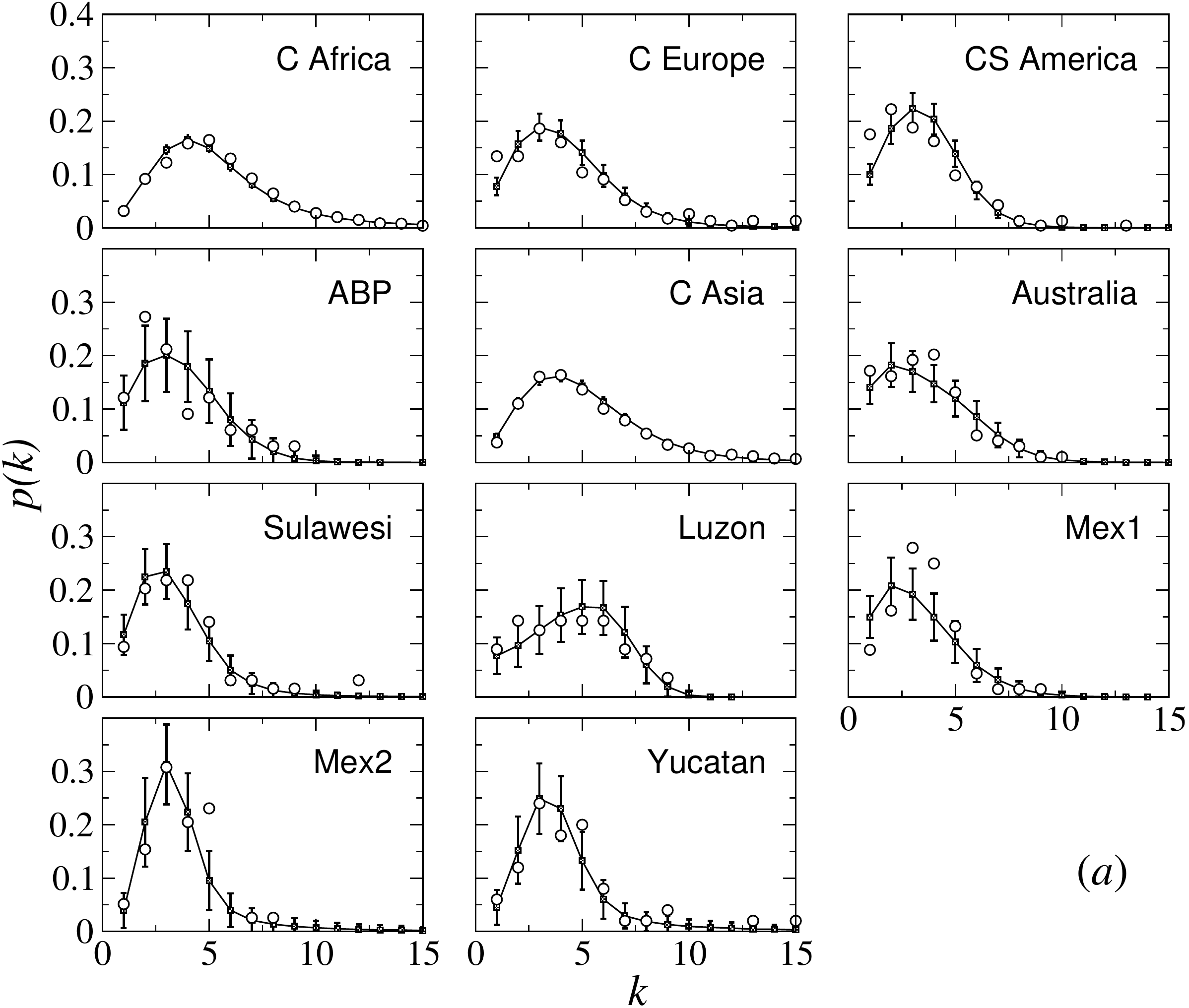}
\includegraphics[width=0.48\textwidth]{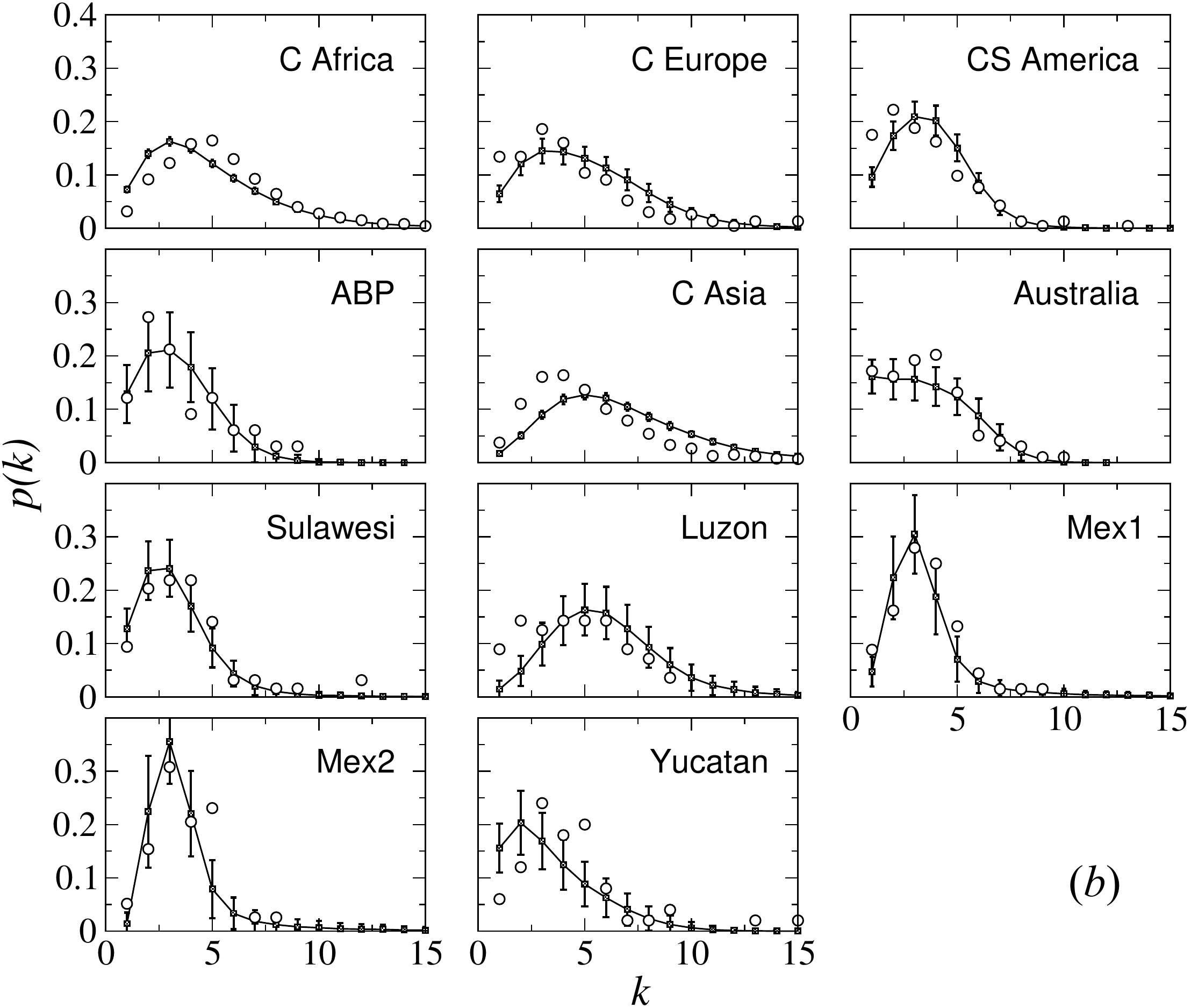}
\caption{\label{fig:ddeg}
Empirical degree distributions (open circles) vs. simulated distributions (linked squares) averaged over $2000$ model 
realizations. Error bars correspond to standard deviations of model-simulated degree distributions. (a) Minimization based 
on the degree distribution. (b) Joint minimization. Ranges of axes are the same in all plots.}
\end{center}
\end{figure}

\begin{figure*}[t!]
\begin{center}
\includegraphics[width=0.49\textwidth]{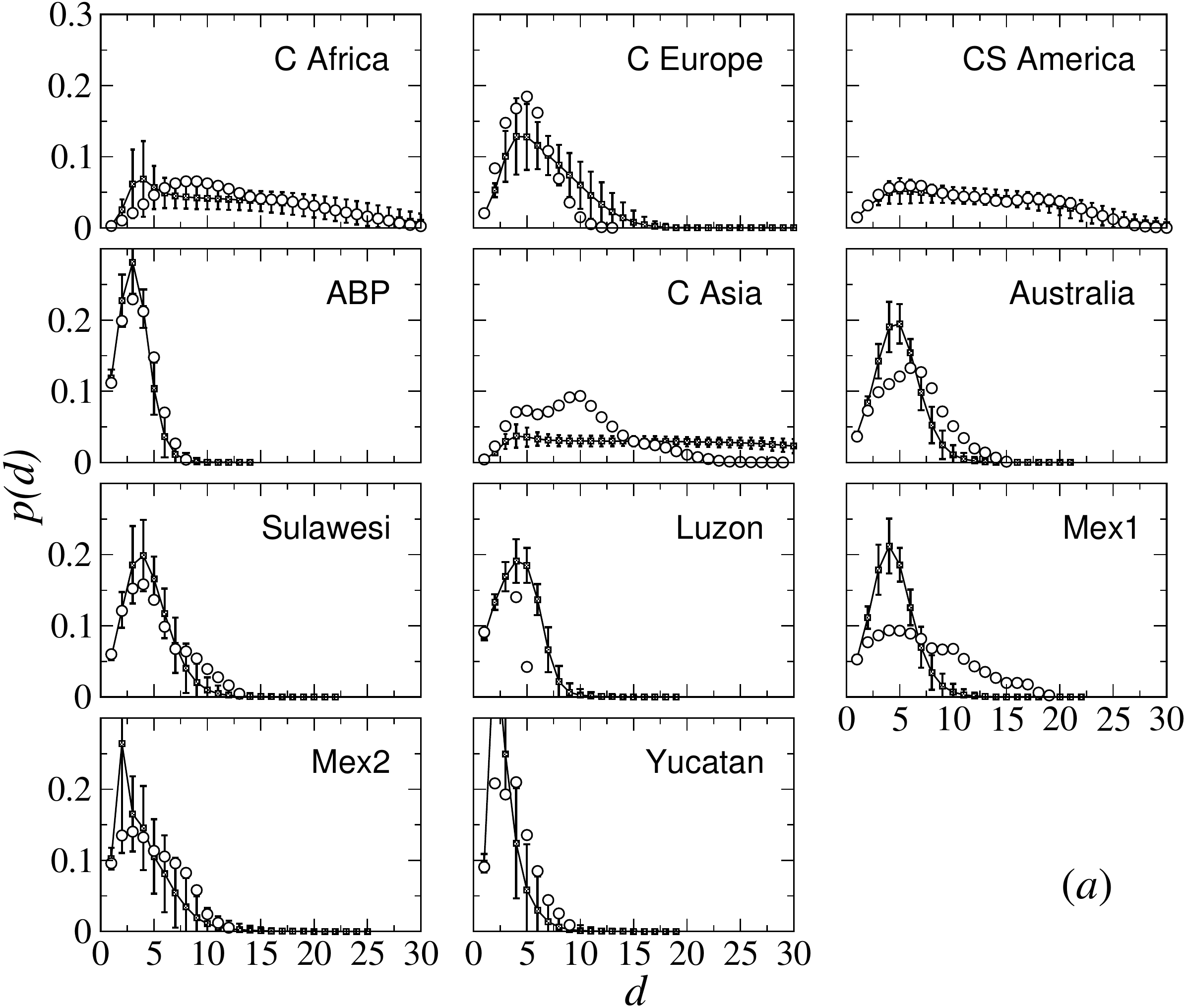}
\includegraphics[width=0.49\textwidth]{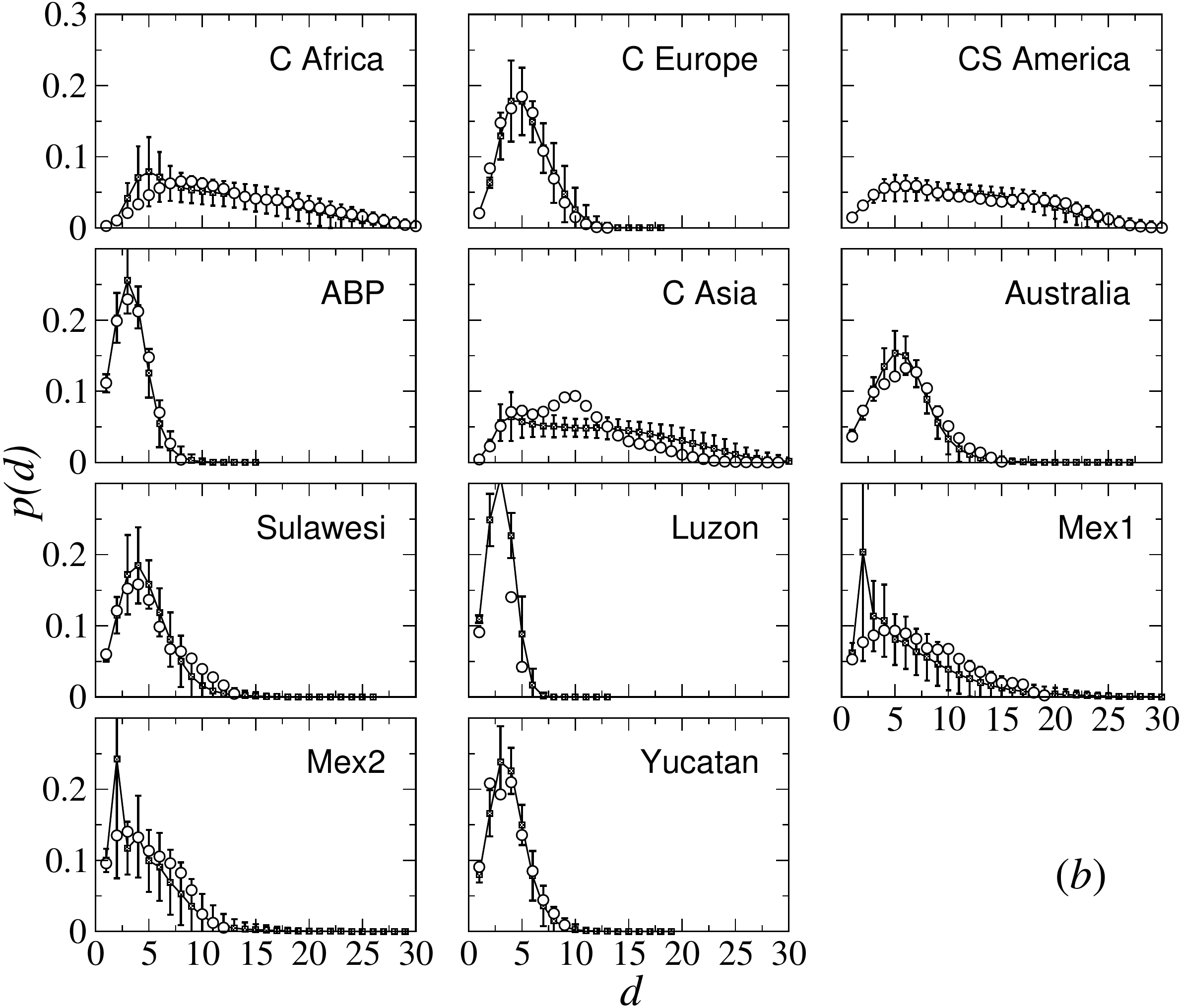}
\caption{\label{fig:dpath}
Empirical shortest-path length distributions (open circles) vs. simulated distributions (linked squares) averaged over 
$2000$ model realizations. Error bars correspond to standard deviations of model-simulated shortest-path length 
distributions. (a) Minimization based on the degree distribution. (b) Joint minimization. Ranges of axes are the same in 
all plots.
}
\end{center}
\end{figure*}

Figure~\ref{fig:ddeg} shows the comparison between empirical and simulated degree distributions. Results in 
Fig.~\ref{fig:ddeg}(a) have been obtained through minimization of Hellinger's distance between degree distributions, 
whereas in Fig.~\ref{fig:ddeg}(b) the result of the joint minimization procedure is shown. In the former case, the 
agreement with empirical data is very good, even when the statistics of the original data (i.e., the network size, see 
Table~\ref{tab:data}) is poor. The Hellinger distances for the degree distribution obtained with the joint minimization 
procedure (cf. Table~\ref{tab:fqfitjoint}) are, as expected, larger than the minimum values reported in 
Table~\ref{tab:fqfitdeg}. 

\subsubsection{Shortest path distributions}

Figure~\ref{fig:dpath} shows the results of the minimization of Hellinger's distance for the degree distribution 
[Fig.~\ref{fig:dpath}(a)] and for the degree and shortest path distribution jointly [Fig.~\ref{fig:dpath}(b)]. In the former 
case, the agreement between empirical and simulated distributions is poor in several cases (and especially in 
continental Asia), but there are some exceptions where the empirical distribution is reasonably reproduced, for example 
in ABP borders, continental South America (the third largest network) or Sulawesi island. The likelihood of the null 
hypothesis that the model can generate networks whose average shortest path length $\langle d \rangle$ is compatible 
with the empirical value $\langle d_{\rm e} \rangle$ has been statistically tested using minimization of Hellinger's distance 
based only on degree distributions. We have calculated the $p$-values of the null hypothesis, 
$\mathrm{Pr}(\langle d\rangle\ge \langle d_{\mathrm{e}}\rangle)$, which are listed in Table~\ref{tab:pvalue}. At a $99\%$ 
confidence level, the null hypothesis is rejected only for continental Asia, Luzon and Mex1. Therefore, even if the 
distribution of shortest-path lengths is not explicitly considered to estimate model parameters, the adaptive network 
model is not statistically rejected to reproduce average path lengths in most empirical networks.

\begin{table}[t!]
\begin{center}
\begin{tabular}{lcc}
\hline\hline
CC & $\langle d_{\mathrm{e}}\rangle$ & $p$-value \\
\hline\hline
C Africa & $12.9$ & $0.46$ \\
\hline
C Asia & $9.5$ & $<10^{-2}$\\
Australia & $6.1$ & $0.02$\\
Sulawesi & $5.2$ & $0.13$\\
Luzon & $2.6$ & $<10^{-2}$\\
\hline
C Europe & $5.0$ & $0.13$\\
\hline
Mex1 & $7.2$ & $<10^{-2}$\\
Yucatan & $3.7$ & $0.14$\\
Mex2 & $4.8$ & $0.18$\\
\hline
CS America & $11.9$ & $0.33$\\
ABP & $3.4$ & $0.17$\\
\hline\hline
\end{tabular}
\end{center}
\caption{
Empirical average path lengths for language networks and $p$-values of the null hypothesis corresponding to the 
adaptive network model with minimization of Hellinger's distance on degree distributions. 
}
\label{tab:pvalue}
\end{table}

The joint minimization significantly improves the fit to empirical shortest-path length distributions, yielding low values of 
Hellinger's distance in most cases, see Table~\ref{tab:fqfitjoint}. Though the joint fit to the degree and the shortest-path 
length distributions worsens the performance of the fit regarding the degree distribution, the overall fit to both distributions 
is significantly improved, as can be seen by comparing the sum $d_{\mathrm H}\text{ (degree)}+d_{\mathrm H}\text{ (path)}$
in Tables~\ref{tab:fqfitdeg} and~\ref{tab:fqfitjoint}.

\subsubsection{Consistency check}

Parameters $r$ and $w$ were obtained at constant values of the perimeter overlap $f$ and the symmetrization 
value $q$, namely $(f^{\star},q^{\star}) = (0.1,0.1)$. To test the consistency of our estimation procedure, we check now 
whether the use of final estimated values in Tables~\ref{tab:fqfitdeg} and~\ref{tab:fqfitjoint} substantially modify the 
performance of the model regarding $z$ and $\rho$. Additional simulations for each $(r,w,f,q)$ set of parameter values 
have been carried out, and averages for exponent $z$ and correlation $\rho$ have been calculated. The results are 
summarized in Table~\ref{tab:avvaldeg}, and should be compared with empirical data in Table~\ref{tab:data}. As can be 
seen, all empirical values lie within the error bars, thus validating {\it a posteriori} the methodology used.

\begin{table}[t!]
\begin{center}
\begin{tabular}{lcccc}
\hline\hline
CC & $z_{\rm s}$ & $\rho_{\rm s}$ & $z_{\rm s}^j$ & $\rho_{\rm s}^j$ \\
\hline\hline
C Africa & $0.91\pm 0.03$ & $0.63\pm 0.01$ & $0.88\pm 0.02$ & $0.63\pm 0.01$ \\
\hline
C Asia & $0.66\pm 0.02$ & $0.72\pm 0.01$ & $0.67\pm 0.02$ & $0.72\pm 0.01$ \\
Australia & $0.67\pm 0.12$ & $0.52\pm 0.08$ & $0.66\pm 0.13$ & $0.52\pm 0.08$ \\
Sulawesi & $0.63\pm 0.07$ & $0.75\pm 0.06$ & $0.63\pm 0.07$ & $0.75\pm 0.06$ \\
Luzon & $0.43\pm 0.05$ & $0.78\pm 0.05$ & $0.44\pm 0.05$ & $0.79\pm 0.05$ \\
\hline
C Europe & $0.59\pm 0.05$ & $0.66\pm 0.04$ & $0.60\pm 0.05$ & $0.65\pm 0.04$ \\
\hline
Mex1 & $0.75\pm 0.14$ & $0.58\pm 0.08$ & $0.87\pm 0.15$ & $0.60\pm 0.08$ \\
Yucatan & $1.34\pm 0.27$ & $0.63\pm 0.09$ & $1.16\pm 0.26$ & $0.60\pm 0.09$ \\
Mex2 & $0.69\pm 0.11$ & $0.74\pm 0.08$ & $0.69\pm 0.11$ & $0.73\pm 0.08$ \\
\hline
CS America & $0.38\pm 0.04$ & $0.56\pm 0.04$ & $0.38\pm 0.04$ & $0.56\pm 0.05$ \\
ABP & $0.64\pm 0.10$ & $0.76\pm 0.08$ & $0.64\pm 0.10$ & $0.75\pm 0.08$ \\
\hline\hline
\end{tabular}
\end{center}
\caption{
For each parameter set obtained through degree distribution minimization, $2000$ model realizations yield the 
estimates $z_{\rm s}$ and $\rho_{\rm s}$, and similarly for the joint minimization. Sub-index ${\rm s}$ stands for 
simulation results; upper-index $j$ indicates joint minimization. Both estimates compare well with the empirical values 
reported in Table~\ref{tab:data}.}
\label{tab:avvaldeg}
\end{table}

\subsubsection{Demographic and topological variables}

\begin{figure}[t!]
\begin{center}
\includegraphics[width=0.48\textwidth]{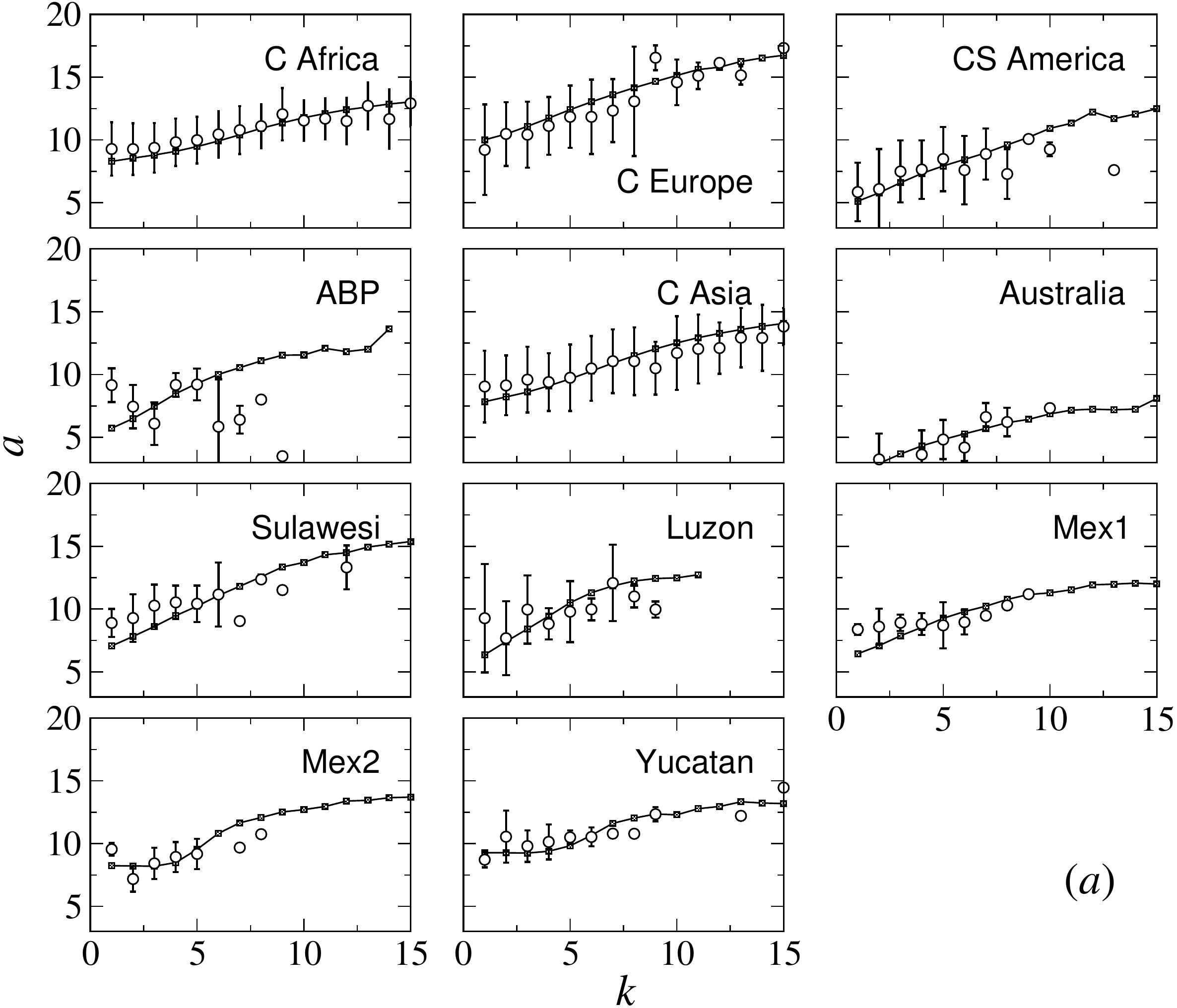}
\includegraphics[width=0.48\textwidth]{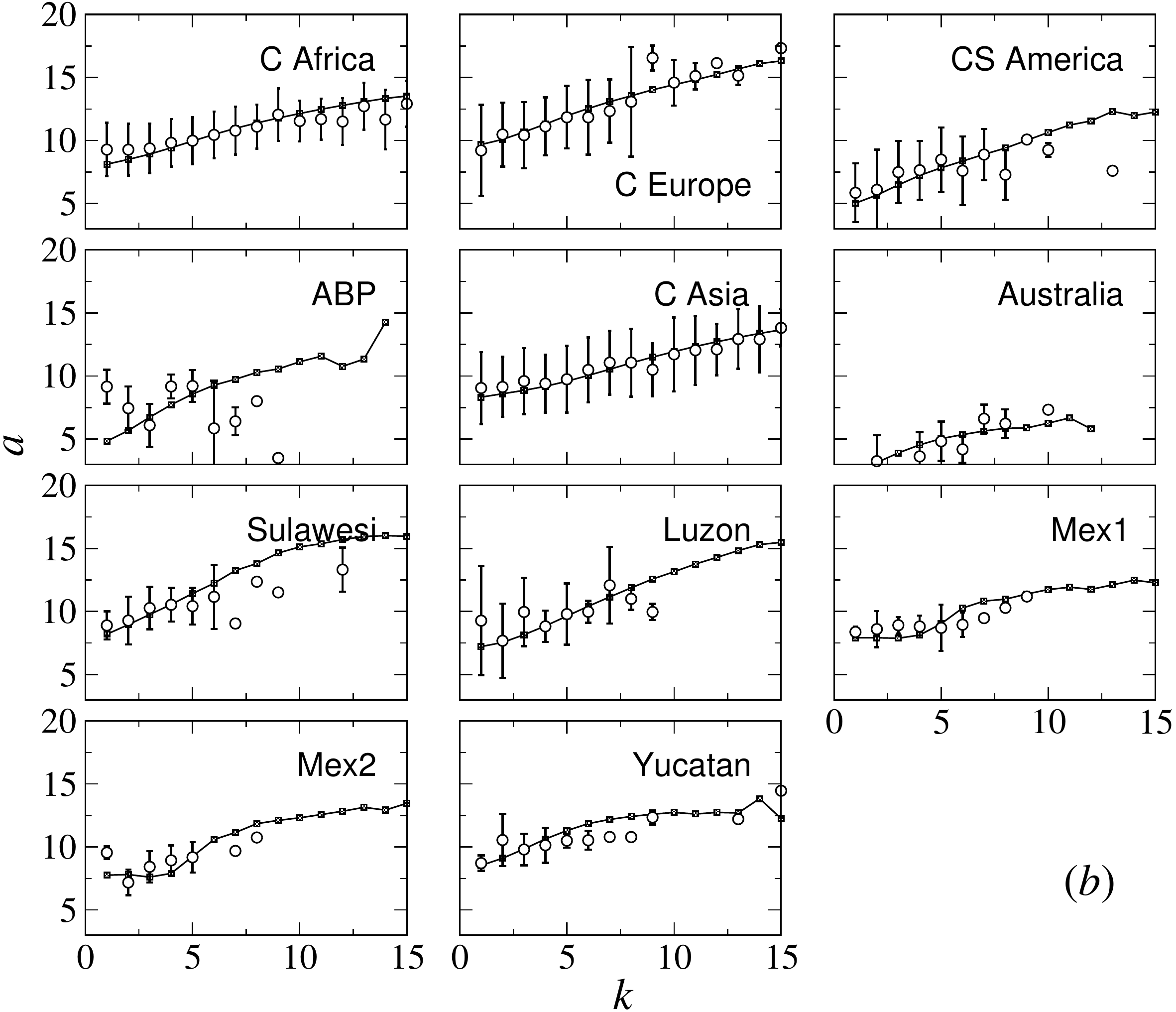}
\caption{\label{fig:ak}
Logarithmic area $a$ vs. node's degree $k$. Black circles correspond to empirical data; error bars are the standard 
deviation of $a$ for each degree. Linked squares are averages over $2000$ realizations of the adaptive network model. 
(a) Minimization based on the degree distribution. (b) Joint minimization. Ranges of axes are the same in all plots. 
}
\end{center}
\end{figure}

\begin{figure*}[t!]
\begin{center}
\includegraphics[width=0.49\textwidth]{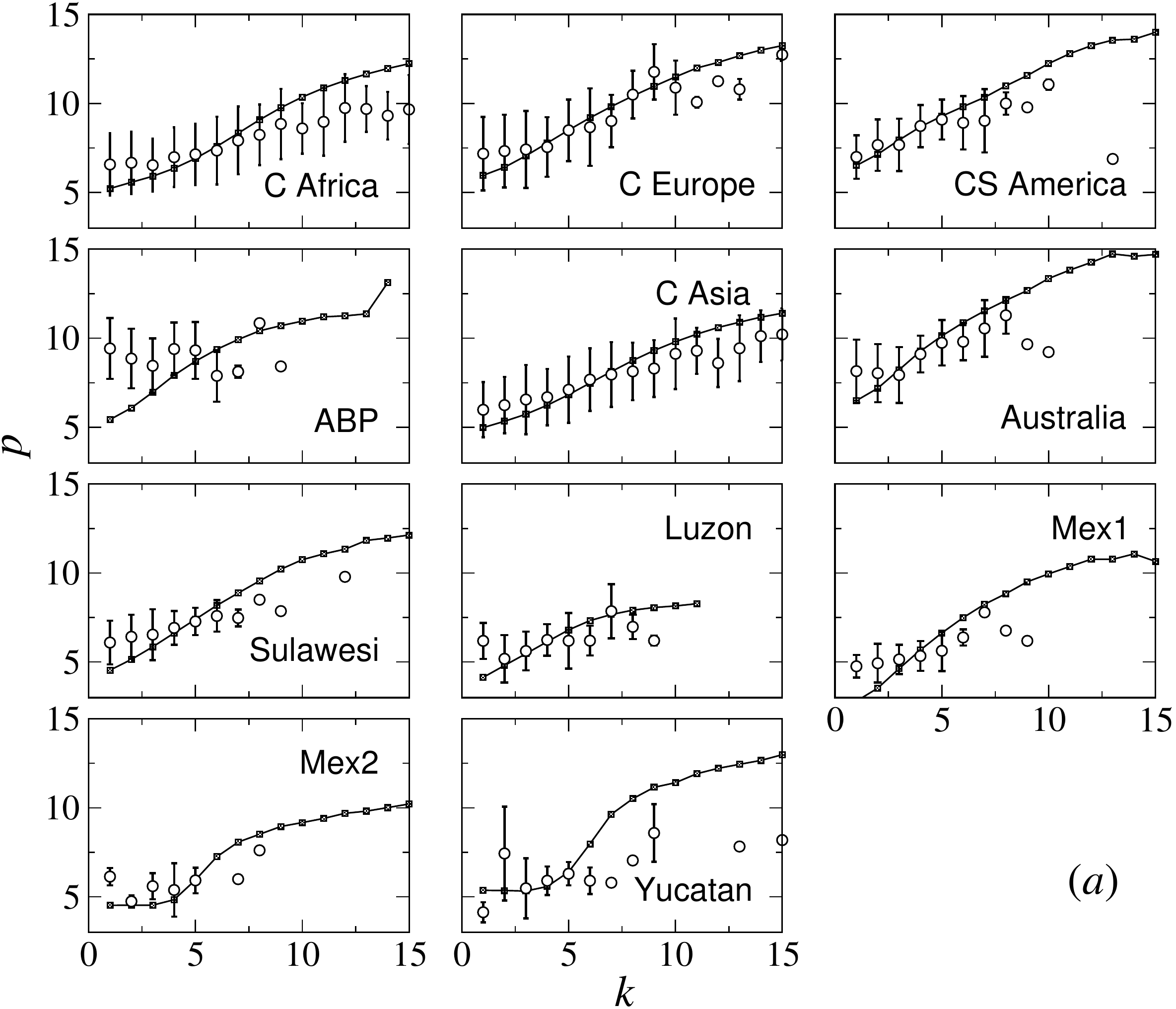}
\includegraphics[width=0.49\textwidth]{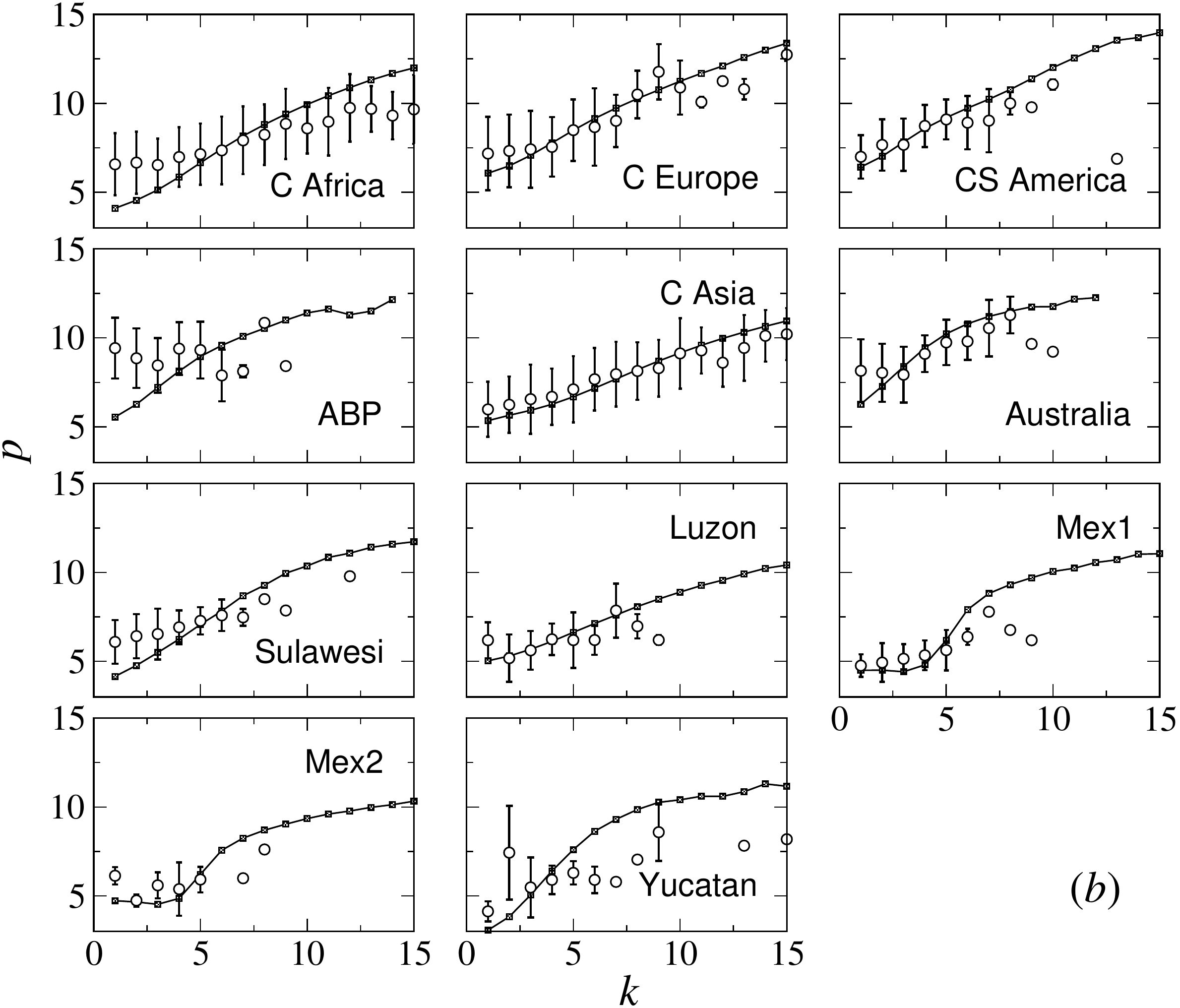}
\caption{\label{fig:pk}
Logarithmic population $p$ vs. node's degree $k$. Black circles correspond to empirical data; error bars are the standard 
deviation of $p$ for each degree. Linked squares are averages over $2000$ realizations of the adaptive network model. 
(a) Minimization based on the degree distribution. (b) Joint minimization. Ranges of axes are the same in all plots. 
}
\end{center}
\end{figure*}

Figures~\ref{fig:ak} and~\ref{fig:pk} depict the correlation between averaged logarithmic areas and averaged 
logarithmic populations, respectively, and degree $k$. Simulation data have been produced with the set of parameters 
obtained under both minimization schemes. For visualization purposes, model results have been displaced in the vertical 
axis through the addition of an arbitrary constant (recall that area and population units are defined up to a constant). Except 
for the ABP network (which is the smallest one, with $N=33$ nodes and therefore a poor statistical power), empirical 
logarithmic areas and populations monotonically increase with $k$. Though these functions are not explicitly considered to 
obtain model parameters, simulations reproduce with remarkable accuracy the empirical observations.

The coupling between the stochastic process that determines population and areas and the update of the network of 
contacts under the perimeter overlap rule leads to the emergence of correlations between the area, population, and degree 
of neighboring nodes. These autocorrelations over the network are measured as
\begin{equation}\label{eq:netcorr}
\tau_i = \frac{1}{2}\sum_{j=1}^N \frac{v_j}{n_{j,i}}\sum_{|k-j|=i}v_k,
\end{equation}
where $v_j$ stands for any of the node properties $a_j$, $p_j$ or $k_j$, and the second sum runs over the nodes which 
are at distance $i$ from node $j$  ($v_j$ are normalized to satisfy $\sum_j v_j=1$ so that autocorrelations for different
variables are independent of their natural scales and can be mutually compared). Distances are measured as 
shortest-path lengths between nodes. We normalize the product $v_jv_k$ by $n_{j,i}$, that is by the number of nodes that 
are at distance $i$ from node $j$. The $1/2$ factor takes into account that all links are double-counted, since the sum 
runs over the whole network. 

\begin{figure}[t!]
\begin{center}
\includegraphics[width=0.49\textwidth]{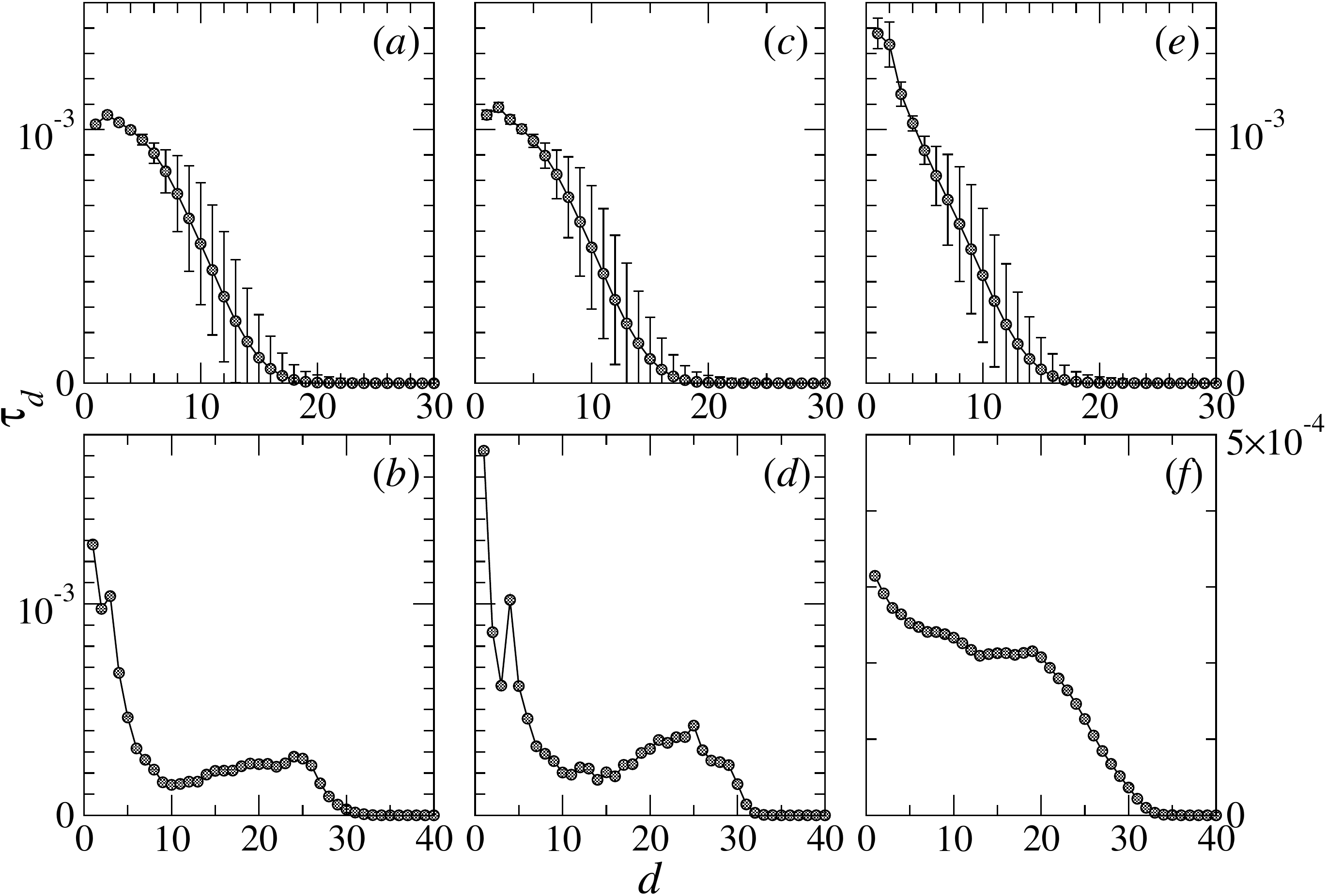}
\caption{\label{fig:netcorr}
Autocorrelation $\tau_d$ over the network, as defined by Eq.~\eqref{eq:netcorr}, as a function of the separation $d$
between nodes, measured in the model
for (a) log-area, (c) log-population, and (e) degree. Model parameters are $r=0.5$, $w=2$, $f=0.1$, and $q=0.1$ for 
networks with $N=500$ nodes. Quantities have been averaged over $500$ independent realizations. For the sake
of comparison, we depict the autocorrelations for (b) log-area, (d) log-population, and (f) degree, obtained for the 
empirical network of continental Africa. Note the different range of the vertical axis in panel (f).
}
\end{center}
\end{figure}

Autocorrelations over the network have been plotted in Figure~\ref{fig:netcorr}. Correlations decay as the separation 
between nodes increases, demonstrating that the network is assortative regarding area, population and node degree. 
These correlations 
cannot be observed if demography and topology are uncoupled: Demographic dynamics without an underlying network of 
contacts correspond to a mean-field model without spatial structure; a static algorithm that reproduces network topology is 
unable to account for correlations in population sizes, a variable disregarded in the algorithm. Model results compare 
only qualitatively with empirical networks (see Fig.~\ref{fig:netcorr} for an example). Following an initial rapid decay,
real networks exhibit an 
intermediate range of node separations where autocorrelations are roughly constant. For large distances, however,
correlations decay as predicted by the mode.

\section{Discussion and conclusions}

The coevolution of population demography and spatial contacts represents a new example of adaptive network in the 
social sciences. By means of a model coupling both processes, we have shown that previous known properties of this 
system are robustly reproduced: population-area relationships and language network topology. Besides, a number of 
features relating demography and topology, as well as certain assortative properties of those networks, are consistently 
obtained in the adaptive network approach here introduced. Assortativity in node population, area and degree are 
by-products induced by subsequent cycles of population change and modification of topological neighborhoods which 
cannot be obtained in scenarios where these two processes are decoupled. Remarkably, the agreement between several 
empirical and simulated quantities is obtained through fits of just four model parameters. We believe this is due to the deep 
meaning of model rules, which are by themselves sufficient to explain the qualitative properties of linguistic groups 
and their associated spatial networks. There is an important exception that cannot be recovered by the model, in particular 
the population-area relationship, which fails to be reproduced already by the mean field approach~\cite{Man12}: New 
Guinea Island. Since this is an often studied example of a region with an extremely high linguistic diversity, it is worth 
mentioning that the demographic and conflict rules we implement do not suffice to yield the $\rho$ value empirically 
measured. In general, the mean-field model is not able to reproduce the set of values $\rho\lesssim 0.5$
for the area-population correlation. Here the set of languages in the giant connected component for New Guinea (note 
that this is a subset of the New Guinean languages used in~\cite{Man12}) yields $\rho=0.42$, which cannot be accounted 
for with the proposed dynamical rules. Similarly, the correlation for all North American languages annotated in the
Ethnologue was not reproduced by the mean-field model proposed in~\cite{Man12}.

The adaptive network model here presented admits a number of extensions. First, the introduction of additional factors 
may make it more realistic. In this study, we have not considered the appearance of new languages or the death of existing 
ones. The origination of new languages can be easily implemented by splitting an existing language. Following our rules, a 
new set of neighbors and an independent evolution of either population appear in a straight way. Death of languages can 
also be considered, for instance, by eliminating those groups whose population falls below a prescribed level (one individual, 
for instance). Quantities such as the average lifetime of languages could be studied in this scenario. Second, model rules 
could be modified to consider ingredients such as language attractiveness or frequency of conflicts dependent on 
degree or population size~\cite{Lim07}. 

Language attractiveness is an important driver in the disappearance of minority languages~\cite{Abr03}, and could be 
implemented through a migration of population from one language to any of its neighbors. In this way, population sizes would 
be modified through processes different from stochastic growth. Extreme versions of migration mechanisms might account 
for the growth of widespread languages~\cite{Dia97}, and perhaps explain the emergence of Dragon Kings in linguistic 
groups~\cite{Sor09}. In the model here used, every group enters into conflict with a neighbor once per time step. This rule 
could be modified to a likely more realistic version where conflict frequency is proportional to the number of links, and not to 
the number of nodes, and the outcome of conflicts could also consider the relative population of the involved parties. 
Also additional cultural markers, as political, linguistic or religious similarities, might modify the frequency and strength of 
conflicts. The results of these modifications are difficult to foresee. Third, the formation of links is now homogeneous and 
does not consider the structure of human settlements in relation to the landscape. The introduction of preferential attachment 
depending on stylized landscape features might help explaining the appearance of a low dimensional niche space in 
language networks~\cite{Cap14,Axe14}. 

The competition for areas between neighbouring populations is a form of demographic conflict. In the scenario here devised, 
these conflicts do not affect population sizes and by definition occur at a characteristic time scale of the order of one year. 
There is a body of literature that has addressed the frequency and distribution of conflicts with the number of casualties in 
terrorist attacks~\cite{Cla07}, wars~\cite{Ced03}, or fatal quarrels in general~\cite{Ric48} as main variable. Those events 
might have frequencies measured in days and have been often modeled as processes of fragmentation and coalescence of 
groups~\cite{Boh09}. It would be interesting to integrate the dynamical network perspective of our study with the fast evolution 
of groups dynamics and its effect on population sizes of these other conflict analyses with the goal of devising more complete 
models for cultural and political clashes. 

Finally, we believe that the model could be applied to other model systems with analogous node and network dynamics. 
One such example is ecology, where an explicit competition for space of species occupying the same niche is known to 
occur. Further, the applicability of the model to that system is supported by a relationship between population sizes and 
ranges of occupation functionally equivalent to the population-area law followed by human linguistic groups. Demographic 
dynamics similar to those used here, perhaps with the addition of temporal biases to grow or decrease, might represent
the dynamics of agents such as companies or religious groups, for example. Suitable modifications of how links are 
established might shed light on the distribution of group sizes and on the relevance of competition and inter-group conflicts. 

\section{Acknowledgments}
The authors are indebted to Jacob B. Axelsen and Dami\'an H. Zanette for their early contribution to some aspects of this 
work. This work was supported by the Spanish MINECO through projects FIS2011-27569, FIS2011-22449 (JAC), and 
CGL2012-39964 (JAC).


\begin{thebibliography}{99}
\bibitem{note} It can be shown that the dispersion in the distribution of language sizes at the beginning of the stochastic 
process (one thousand years ago) has a negligible contribution to the value of $\sigma_{\alpha}$ (at most 0.5\%, obtained 
from the limit situation where each language had one single speaker). The figure here reported stems thus from a 
multiplicative stochastic process in population growth with arbitrary initial conditions, average value 1.0029, and a mean 
square dispersion caused by that of the current distribution of extant languages. See [19] for a full derivation of parameters 
$\alpha$ and $\sigma_{\alpha}$.
\bibitem{note1} Note that the distribution of $\beta_i$ is related to the distribution of the (untransformed) growth rate
$\alpha_i$ as $P(\beta) d\beta = Q(\alpha) d\alpha$, with $\beta_i = \log \alpha_i$. Then $\epsilon = \log \alpha$ and $\eta$ 
is derived from the relation between the two distributions.
\end{thebibliography}
\end{document}